\documentclass[12pt]{iopart}

\begin{document}

\title[]{Quantum gravity correction to Hawking radiation of the 2+1 dimensional wormhole}

\author{Ganim Gecim$^{1}$ and Yusuf Sucu$^{2}$}

\address{$^{1}$Department of Astronomy and Astrophysics, Faculty of Science, Atat\"{u}rk Univ.,25240 Erzurum, Turkey}
\address{$^{2}$Department of Physics, Faculty of Science, Akdeniz University, 07058 Antalya, Turkey}
\ead{ggecim@atauni.edu.tr and ysucu@akdeniz.edu.tr}

\vspace{10pt}

\begin{abstract}
We carry out the Hawking temperature of a (2+1) dimensional circularly symmetric traversable wormhole in the framework of the generalized uncertainty principle (GUP). Firstly, we introduce the modified Klein-Gordon equation of the spin-0 particle, the modified Dirac equation of the spin-1/2 particle, and the modified vector boson equation of the spin-1 particle in the wormhole background, respectively. Given these equations under the Hamilton-Jacobi approach, we analyze the GUP effect on the tunneling probability of these particles near the trapping horizon, and subsequently, on the Hawking temperature of the wormhole. Furthermore, we have found that the modified Hawking temperature of the wormhole is determined by both wormhole's and tunneling particle's properties and indicated that the wormhole has a positive temperature similar to that of a physical system. This case indicates that the wormhole may be supported by ordinary (non-exotic) matter. In addition, we calculate the Unruh-Verlinde temperature of the wormhole by using Kodama vectors instead of time-like Killing vectors, and observe that it equals to the standard Hawking temperature of the wormhole.
\end{abstract}

\section{Introduction} \label{intro}

Black hole and wormhole solutions are the most popular and fascinating solutions of Einstein general relativity as well as modified gravitational theories. Theoretically, the black hole consists of a singularity in the center and an event horizon that surrounds this singularity, but the wormhole consists of two mouths and a throat connecting them. Also, the mouths of a wormhole open into two different regions of the same space-time or different two space-times in the presence of an exotic matter called phantom violating the energy conditions \cite{Morris1,Morris2}. However, it has been shown that a wormhole can be formed without an exotic matter, i.e., it can be supported by ordinary matter not violating energy condition \cite{Rich,Band,Mazhar1,Mazhar2,Mazhar3,Mazhar4,Kanti,Harko1,Samanta,Myrzakulov}. Furthermore, a wormhole space-time is topologically similar to a black hole except for the nature of their horizon. This situation arise from the fact that the wormhole horizon is characterized by a \textit{temporal outer trapping horizon} while the black hole horizon is characterized by a \textit{future spatial trapping region} \cite{Hayward1}. Thus, thanks to the its throat, it is believed that a wormhole allows a traveler to both directions.

After the seminal papers of Hawking \cite{Haw1,Haw2,Haw3}, the thermodynamic properties of the these cosmological objects have been a great interest. From these thermodynamical properties, especially, thermal radiation of a black hole, named as Hawking radiation, is calculated by various methods. From these methods, the Hamilton-Jacobi method, based on quantum tunneling process of a particle, is the most popular method \cite{Kraus1,Kraus2,Parikh1,Parikh2,Ang,Srin,Shan,Kerner1,Kerner2,Kerner3,Kerner4}. By using the Hamilton-Jacobi approach, Hawking radiations of a number of black holes as quantum tunneling process of a point particle have been investigated in the literature \cite{Jusufi,GY1,GY2,GY3,Li1,Li2,Ejaz,Chen1,VEC1,VEC2,VEC3,Rocha,Caval}. Recently, it has been also showed that Elko (dark) particles which are spin-1/2 fermions of mass dimension one and dual-helicity eigen spinors of the charge conjugation operator are emitted at the expected Hawking temperature from black holes and strings \cite{Rocha,Caval}. These studies show that Hawking radiation depends only on the black hole's properties. On the other hand, when the thermodynamic properties of a black hole are examined in the context of the GUP which is based on the existence of a minimal observable length that is a characteristic of the candidate theories of quantum gravity it is seen that the modified Hawking temperature is up to the properties of both the black hole and the tunnelled particle \cite{noz1,noz2,GWS1,GWS2,GWS3,a1,a2,a3,a4,a5,a6,a7,a8,a9,a10,a11,a12,a13,Haldar,Sha,Kuntz,Casadio}. Furthermore, in the presence of the GUP effect, it was observed that the particles with spin (spin-0,1/2,1) are differently tunnelled from a black hole, and therefore, they caused completely different Hawking temperature \cite{GY4,GY5,GY6,GY7,GY8,GY9}. However, while the black hole thermodynamic is intensively studied in the literature there are few studies deal with thermodynamic properties of wormholes \cite{Hotta,Diaz1,Diaz2,Diaz3,Diaz4,Saiedi,Umar,Debnath1,Debnath2,Jamil}. The existence of a trapped horizon would allow the use of semi-classical approaches  such as Hamilton-Jacobi approach based on the tunneling of particles for investigation of the thermodynamic properties of a wormhole. Thus, studies investigating the temperature of a wormhole by quantum mechanical tunneling process of the spinning particles show that the temperature of the wormholes is not related to the properties of the tunneling particles \cite{Sharif,Sakalli1,Sakalli2}. Moreover, according to our knowledge, the thermodynamic properties of wormholes have not yet been analyzed in the context of quantum gravity. Therefore, we will look into the Hawking temperature of the (2+1) dimensional static traversable wormhole via quantum mechanical tunnelling of the massive spin-0, spin-1/2 and spin-1 particles in the context of GUP.

The purpose of the paper is investigate the modified Hawking temperature of the (2+1) dimensional wormhole by using the Hamilton-Jacobi approach in the context of the GUP. Hence, in the following section, using the modified Klein-Gordon equation, we calculate the tunneling probability of the scalar particle, and subsequently, the modified Hawking temperature of the wormhole. In this section, we also derive the Unruh-Verlinde temperature by using the Kodama vector instead of time-like Killing vector, and compare it with the standard Hawking temperature derived by quantum tunnelling process. In section-3 and 4, for the modified Dirac equation that describes spin-1/2 Dirac particle and for the modified vector boson equation that describes spin-1 vector boson particle, respectively, we obtain the modified temperature of the wormhole by the same procedure for the scalar particle. In the conclusion, we are summarized the results.

\section{Hawking temperature of the 2+1 dimensional traversable wormhole}\label{worm}

Despite of significant studies, the quantum nature of gravity continues to hide its mystery, and this is still one of the most important problems of theoretical physics. Recently, the interest in developing theories for understanding the nature of gravity in lower dimensions is growing. Because, lower dimensional theories are both mathematically simple, and help to stimulate new insights into its higher dimensional counterparts by generating new ideas. In particular, several (2+1) dimensional gravity models have been built up that guide to the understanding of the problems encountered in their (3+1) dimensional counterparts \cite{Deser1,Deser2,Witten1,Witten2,Berg,Clement1,Clement2}. On the other hand, in the Planck scale, a minimal length appears naturally in the suitable candidate theories for the quantum gravity such as string theory, loop quantum gravity theory and noncommutative geometry \cite{Hin,Kempf,Afa,Sdas,Carr}. Furthermore, to analyze the effect of the quantum gravity on a curved spacetime background, (2+1) dimension satisfies a simplicity with respect to (3+1) dimension.

The general form of the (2+1) dimensional circularly symmetric traversable wormhole spacetime is given as follows \cite{Perry,Mazhar,Raha,Kim};
\begin{eqnarray}
ds^{2}=e^{2f(r)}dt^{2}-\frac{1}{1-\frac{b(r)}{r}}dr^{2}-r^{2}d\phi^{2}, \label{metric1}
\end{eqnarray}
where the $b(r)$ and $f(r)$ are the shape and the red-shift functions, respectively. To represent a wormhole via Eq.(\ref{metric1}), there are some constraints known as flare-out conditions on the $f(r)$ and $b(r)$ functions: To guarantee the absence of a horizon, the red-shift function, $f(r)$, must be taken finite values everywhere. At throat, i.e. $r$=$r_{th}$ where $r_{th}$ is the radius of the throat, the shape function, $b(r)$, must be obeyed the conditions $b(r_{th})$=$r_{th}$ and $b^{\prime}(r_{th})<1$. Also, the shape function must be obeyed $b(r)<r$ for $r>r_{th}$ and $b(r)/r\rightarrow 0$ in the limit $\left\vert r\right\vert \rightarrow \infty$.

In this study, we will use the Kodama vector, $K^{a}$, to calculate the surface gravity \cite{Kodama,Cris1,Cris2}. The Kodama vector lies in the $(t-r)$ plane and its definition is given as follows:
\begin{eqnarray}
K^{a}=\frac{1}{\sqrt h}\varepsilon^{ab}\partial_{b}R \label{kodama}
\end{eqnarray}
where $h$ is the determinant of the metric tensor $h_{ab}$ of the (1+1)-dimensional plane, $\varepsilon^{ab}$ is the (1+1)-dimensional Levi-Civita tensor, and $R$ is radial function \cite{Kodama,Cris1,Cris2}. For this purpose, the (2+1) dimensional traversable wormhole metric Eq.(\ref{metric1}) can be rewritten as follows:
\begin{eqnarray}
ds^{2}=h_{ab}dx^{a}dx^{b}-r^{2}d\phi^{2}, \ \ (a,b=0,1) \label{metric2}
\end{eqnarray}
where $x^{a}$=$(t,r)$ and $h_{ab}$=$(1,-\frac{r}{r-b(r)})$ is the metric of the two dimensional space. Here, we consider the zero-redshift case, i.e. $f(r)$=$0$. Thus, the components of the Kodama vector for the (2+1) dimensional traversable wormhole metric Eq.(\ref{metric2}) as follows:
\begin{eqnarray}
K^{a}=\left(\sqrt{1-\frac{b(r)}{r}},0,0\right). \label{kodama2}
\end{eqnarray}
Furthermore, using the Kodama vector, the energy of a particle can be defined as follows:
\begin{eqnarray}
E=-K^{a}\partial_{a}S, \label{kodamaE}
\end{eqnarray}
where $S$ is the action function of a particle \cite{Kodama,Cris1,Cris2}. On the other hand, the radius of trapping horizon is determined by the following way \cite{Jamil,Cai,Li}
\begin{eqnarray}
h^{ab}\partial_{a} r \partial_{b} r=0. \label{metric3}
\end{eqnarray}
Hence, using Eq.(\ref{metric3}), we calculate the radius of the trapping surface as $r_{trap}$=$b(r_{trap})$  means that the radius of trapping horizon is equal to radius of the throat, i.e. $r_{trap}$=$r_{th}$.

This result exhibit that radius of the trapping horizon of static traversable wormhole overlap with radius of its throat. Using this fact, we aim to investigate the quantum mechanical effects near the trapping horizon by using Hamilton-Jacobi approach in the context of GUP. To do so, we use the modified Dirac, Klein-Gordon and vector boson equations.

\subsection{Tunneling of scalar particle from the wormhole}\label{Scalar}

To clarify the quantum gravity correction to the tunneling of the scalar particle from the wormhole, we start by writing the modified Klein-Gordon equation given as follows:
\begin{eqnarray}
\hbar^{2}\partial_{t}\partial^{t}\widetilde{\Phi}+\hbar^{2}\partial_{i}\partial^{i}\widetilde{\Phi}+ 2\alpha \hbar^{4}\partial_{i}\partial^{i}(\partial_{i}\partial^{i}\widetilde{\Phi})+M_{0}^2(1-2\alpha M_{0}^2)\widetilde{\Phi}=0, \label{ModKG}
\end{eqnarray}
where $\widetilde{\Phi }$ and $M_{0}$ are the modified wave function and mass of the scalar particle, respectively, and also the GUP parameter, $\alpha$, is defined as $\alpha$=$\alpha_{0}/M_{p}^{2}$ in terms of $M_{p}$ (the Planck mass) and a dimensionless
parameter, $\alpha_{0}$. The upper limit for $\alpha_{0}$ is reported as $10^{21}$ \cite{Das,Scard,Ghosh}. By using the wormhole background in Eq.(\ref{metric1}), the modified Klein-Gordon equation is rewritten in the following form:
\begin{eqnarray}
\hbar^{2}\frac{\partial^{2}\widetilde{\Phi}}{\partial t^{2}}-\frac{\hbar^{2}}{r^{2}}
\frac{\partial ^{2}\widetilde{\Phi }}{\partial \phi^{2}}+2\alpha \hbar^{4}C(r)\frac{\partial^{2}}{\partial r^{2}}\left(C(r) \frac{\partial^{2}\widetilde{\Phi }}{\partial r^{2}}\right)
+\frac{2\alpha \hbar^{4}}{r^{2}}\frac{\partial ^{2}}{\partial \phi^{2}}\left(\frac{1}{r^2}\frac{\partial^{2}
\widetilde{\Phi }}{\partial \phi^{2}}\right) \nonumber \\ -\hbar^{2}C(r)\frac{\partial^{2}\widetilde{\Phi}}{\partial r^{2}}+M_{0}^{2}\left(1-2\alpha M_{0}^{2}\right)\widetilde{\Phi}=0,
\label{ScalarD}
\end{eqnarray}
where $C(r)$=$1-\frac{b(r)}{r}$. Furthermore, the modified wave function of the scalar particle, $\widetilde{\Phi}\left(t,r,\phi \right)$, is defined as
\begin{eqnarray}
\widetilde{\Phi}\left(t,r,\phi \right)=A\exp \left(\frac{i}{\hbar }S\left( t,r,\phi \right) \right), \label{ansatzKG}
\end{eqnarray}
where $A$ is a constant and $S(t,r,\phi )$ is the classical action function. When inserting Eq.(\ref{ansatzKG}) into Eq.(\ref{ScalarD}), we get the modified Hamilton-Jacobi equation as follows:
\begin{eqnarray}
\left(\frac{\partial S}{\partial t}\right)^{2}-C(r)\left(
\frac{\partial S}{\partial r}\right)^{2}-\frac{1}{r^{2}}\left(\frac{\partial S}{\partial\phi}\right)^{2}-M_{0}^{2}-\frac{2\alpha}{r^{4}}\left(\frac{\partial S}{\partial \phi}\right)^{4} \nonumber
\\ +2\alpha\left[M_{0}^{4}-C(r)^2\left(\frac{\partial S}{\partial r}\right)^{4}\right]=0, \label{HamiltonKG}
\end{eqnarray}
where the higher order terms of $\hbar$ are neglected. The action function, $S\left(t,r,\phi \right)$, can be separated by means of the Kodama vector \cite{Kodama,Cris1,Cris2} which the Kodama vector for the wormhole background is $K^{i}$=$(\sqrt{C},0,0)$ and also the energy of the tunnelling particle, $E$, is described by the Kodama vector as $E$=$-\sqrt{C}\frac{\partial S}{\partial t}$. Accordingly, by using the separation of variable method the action is written by the following form:
\begin{eqnarray}
S\left(t,r,\phi \right)=-\int{\frac{E}{\sqrt{C(r)}}}dt+j\phi +W(r)+k \label{Radial}
\end{eqnarray}
where $k$ is a complex constant, and $j$ is the angular momentum of the particle. In this context, the modified Hamilton-Jacobi equation is reduce to the radial trajectory, $W(r)$:
\begin{equation}
W_{\pm}(r)=\pm \int \frac{\sqrt{E^2-C(r)\left(M_{0}^{2}+j^{2}/r^{2}\right)}}{C(r)}\left[1+\alpha \Lambda\right]dr,
\label{radialKG1}
\end{equation}
where the signs $(+/-)$ are represent to the outgoing particle and ingoing particle, respectively. Also, the abbreviation $\Lambda$ is
\begin{equation*}
\Lambda=\frac{C(r)^{2}\left(M_{0}^{4}-j^{4}/r^{4}\right)-\left[E^{2}-C(r)\left(M_{0}^{2}+j^{2}/r^{2}\right)\right]^{2}}{C(r)\left[E^{2}-C(r)\left(M_{0}^{2}+j^{2}/r^{2}\right)\right]}.
\end{equation*}
In the near trapping horizon limit, $C(r)\approx(r-r_{h})a$ where $a$ is a constant and it is
\begin{eqnarray}
a=\frac{b(r_{h})}{r_{h}^{2}}-\frac{b^{\prime}(r_{h})}{r_{h}}, \label{constant}
\end{eqnarray}
where the prime denote the derivative with respect to $r$. Because of the flare out conditions, we see that the constant $a$ should be positive, i.e. $a>0$. In that case, the $W_{\pm}(r)$ are calculated as
\begin{equation}
W_{\pm}(r_{h})=\pm i\pi\frac{E}{a}\left[1+\frac{3\alpha}{2}(M_{0}^{2}+\frac{j^{2}}{r_{h}^{2}})\right],\label{radialKG2}
\end{equation}
Then, the scalar particle tunneling probabilities from the trapping horizon are
\begin{eqnarray}
P_{out}=\exp \left[-\frac{2}{\hbar}ImW_{+}\left(r\right)\right],
\nonumber \\
P_{in}=\exp \left[-\frac{2}{\hbar}ImW_{-}\left(r\right)
\right]. \label{Eqn1}
\end{eqnarray}
Thus, the tunneling probability is
\begin{equation}
\Gamma=e^{-\frac{2}{\hbar}ImS}=\frac{P_{out}}{P_{in}}=e^{-\frac{E}{T_{H}}},\label{Eqn2}
\end{equation}
where $T_{H}$ is Hawking temperature. Consequently, the modified Hawking temperature of the scalar particle, $T_{H}^{KG}$, is obtained as
follows
\begin{eqnarray}
T_{H}^{KG}=\frac{T_{H}}{\left[1+\frac{3\alpha}{2}(M_{0}^{2}+\frac{j^{2}}{r_{h}^{2}})\right]}\approx T_{H}\left[1-\frac{3\alpha}{2}(M_{0}^{2}+\frac{j^{2}}{r_{h}^{2}})\right]\label{Eqn4}
\end{eqnarray}
where $T_{H}$ is the usual Hawking temperature of the wormhole and it is explicitly given by
\begin{eqnarray}
T_{H}=\frac{\hbar a}{4\pi}. \label{Eqn5}
\end{eqnarray}
Since $a$ is positive, the usual Hawking temperature of the wormhole, $T_{H}$, also positive. This result shows that we can take the wormhole as a physical system, i.e. it may supported by an ordinary (non-exotic) matter. In the previous studies, it was emphasized that Hawking temperature of a wormhole should be negative due to the exotic matter that supports it. In these studies the trapping horizon of the wormhole is consider in the \textit{past outer trapping region} \cite{Diaz4,Sakalli1,Sakalli2}. However, our result indicates that the wormhole may supported by ordinary (non-exotic) matter. In addition, our result shows that the trapping horizon of the wormhole is represented by a \textit{future outer trapping region} \cite{Hayward2,Hayward3}. Moreover, according to Eq.(\ref{Eqn4}), we can say that the standard Hawking temperature is higher than the modified Hawking temperature. In that case, the mass and angular momentum of the tunneling scalar particle play an important role in determining the thermodynamic properties of the wormhole, as in scalar particle tunneling from both (2+1) and (3+1) dimensional black holes in the presence of the GUP effect \cite{a4,a7,a9,a10,GY4,GY6,GY7,GY8,GY9}.

In this point, it is important to emphasize that the standard Hawking temperature of the wormhole given in Eq.(\ref{Eqn5}) can be derived in the context of Unruh-Verlinde approach \cite{Unruh,Verlinde}. In the standard Verlinde formulation, Unruh-Verlinde temperature defined in the following expression \cite{Verlinde,Konoplya,Liu}:
\begin{eqnarray}
T_{Unruh}=\frac{\hbar}{2\pi}e^{\Phi}n^{\alpha}\nabla_{\alpha}\Phi, \label{UV1}
\end{eqnarray}
where $n^{\alpha}$ is unit vector and $\Phi$ is the modified Newtonian potential given in terms of metric tensor, $g^{\mu\nu}$, and time-like Killing vector, $\xi_{\mu}$, as \cite{Wald}
\begin{eqnarray}
\Phi=\frac{1}{2}\ln(-g^{\mu\nu}\xi_{\mu}\xi_{\nu}). \label{UV2}
\end{eqnarray}
As indicated in \cite{Konoplya}, the Unruh-Verlinde temperature calculated using Eq.(\ref{UV1}) is zero. However, if we use the Kodama vector instead of the time-like Killing vector in Eqs.(\ref{UV1}) and (\ref{UV2}), we find a non-zero Unruh-Verlinde temperature. Thus, we obtain the Unruh-Verlinde temperature of the (2+1) dimensional traversable wormhole as follows:
\begin{eqnarray}
T_{Unruh}=\frac{\hbar}{4\pi}\left(\frac{d C(r)}{dr}\right)_{r=r_{h}}=\frac{\hbar}{4\pi}\left(\frac{b(r_{h})}{r_{h}^{2}}-\frac{b^{\prime}(r_{h})}{r_{h}}\right) \label{UV3}
\end{eqnarray}
which is consistent with Eq.(\ref{Eqn5}). According to this result, both standard Hawking temperature and the Unruh-Verlinde temperature are non-vanishing at throat of the 2+1 dimensional traversable wormhole and are equal to each other. Furthermore, this situation shows that a \textit{Kodama observer} measures a different temperature value than that of a \textit{Killing observer}.

\subsection{Tunneling of massive Dirac particle from the wormhole}\label{Dirac}

The modified (2+1) dimensional Dirac equation by using the GUP relations is given by the following way;
\begin{eqnarray}
i\overline{\sigma}^{0}(x)\partial_{0}\widetilde{\Psi}+i\overline{\sigma
}^{i}(x)\left(1-\alpha\mu_{0}^{2}\right)\partial_{i}\widetilde{\Psi}-\frac{\mu_{0}}{\hbar}\left(1+\alpha\hbar^{2}\partial_{j}\partial^{j}-\alpha \mu_{0}^{2}\right)\widetilde{\Psi} \nonumber \\+i\alpha\hbar^{2}\overline{\sigma}^{i}(x)\partial_{i}\left(\partial_{j}\partial^{j}\widetilde{\Psi}\right)-i\overline{\sigma}^{\mu}(x)\Gamma_{\mu}\left(1+\alpha\hbar^{2}\partial_{j}\partial^{j}-\alpha\mu_{0}^{2}\right)\widetilde{\Psi}=0,\label{ModD}
\end{eqnarray}
where $\widetilde{\Psi}$ is the modified Dirac spinor, $\mu_{0}$ is mass of
the Dirac particle, $\overline{\sigma}^{\mu}(x)$ are the spacetime
dependent Dirac matrices, and $\Gamma_{\mu}(x)$ are spin affine connection
for spin-1/2 particle \cite{GY4,GY6,GY7,GY8,Sucu1}. Using Eq.(\ref{metric1}), the non-zero spinorial
affine connection is derived in terms of Pauli matrices as follows;
\begin{eqnarray}
\Gamma_{2}=\frac{1}{2}\sqrt{C(r)}{\sigma}^{1}{\sigma}^{2}. \label{connec}
\end{eqnarray}
To get the tunneling probability of the massive Dirac particle from the wormhole, we use the following ansatz;
\begin{equation}
\widetilde{\Psi}(x)=\exp\left(\frac{i}{\hbar}S\left(t,r,\phi\right)\right)\ \left(\begin{array}{c}A\left(t,r,\phi\right)
\\B\left( t,r,\phi \right) \\ \end{array}\right) \label{ansatzD}
\end{equation}
where the $A\left(t,r,\phi\right)$ and $B\left(t,r,\phi\right)$ are the functions of spacetime coordinates \cite{Sucu1}. Substituting the Eqs.(\ref{connec}) and (\ref{ansatzD}) into the Eq.(\ref{ModD}), we obtain the following coupled
equations for the leading order in $\hbar$ and $\alpha$:
\begin{eqnarray*}
B\left[i\sqrt{C(r)}\left( 1-\alpha \mu_{0}^{2}\right)\frac{\partial S}{\partial r}+\frac{\left(1-\alpha\mu_{0}^{2}\right)}{r}\frac{\partial S}{\partial\phi}+i\alpha C(r)^{3/2}\left(\frac{\partial S}{\partial r}\right)^{3}\right]\\
+B\left[i\frac{\alpha \sqrt{C(r)}}{r^{2}}\frac{\partial S}{\partial r}\left(\frac{\partial S}{\partial \phi}\right)^{2}+\frac{\alpha C(r)}{r}\frac{\partial S}{\partial \phi}\left(\frac{\partial S}{\partial r}\right)^{2}+
\frac{\alpha}{r^{3}}\left(\frac{\partial S}{\partial \phi}\right)^{3}\right] \\ +A\left[\frac{\partial S}{\partial t}+\mu_{0}\left(1-\alpha\mu_{0}^{2}\right)+\frac{\alpha\mu_{0}}{r^{2}}\left(\frac{\partial S}{\partial \phi}\right)^{2}+\alpha\mu_{0}C(r)\left(\frac{\partial S}{\partial r}\right)^{2}\right]=0,
\end{eqnarray*}
\begin{eqnarray}
A\left[-i\sqrt{C(r)}\left(1-\alpha\mu_{0}^{2}\right)\frac{\partial S}{\partial r}+\frac{\left(1-\alpha\mu_{0}^{2}\right)}{r}\frac{\partial S}{\partial\phi}-i\alpha C(r)^{3/2}\left(\frac{\partial S}{\partial r}\right)^{3}\right] \nonumber \\
+A\left[-i\frac{\alpha \sqrt{C(r)}}{r^{2}}\frac{\partial S}{\partial r}\left(\frac{\partial S}{\partial \phi}\right)^{2}+\frac{\alpha C(r)}{r}\frac{\partial S}{\partial\phi}\left(\frac{\partial S}{\partial r}\right)^{2}+
\frac{\alpha}{r^{3}}\left( \frac{\partial S}{\partial \phi}\right)^{3}\right] \nonumber \\
+B\left[\frac{\partial S}{\partial t}-\mu_{0}\left(1-\alpha \mu_{0}^{2}\right)-\frac{\alpha \mu_{0}}{r^{2}}\left(\frac{\partial S}{\partial \phi}\right)^{2}-\alpha \mu_{0}C(r)\left(\frac{\partial S}{\partial r}\right)^{2}\right]=0. \label{EqnD1}
\end{eqnarray}
Thus, we obtain the modified Hamilton-Jacobi equation describing the massive Dirac particle for vanishing determinant of the coefficient matrix of $A\left(t,r,\phi\right)$ and $B\left(t,r,\phi\right)$:
\begin{eqnarray}
\left(\frac{\partial S}{\partial t}\right)^{2}-C(r)\left(\frac{\partial S}{\partial r}\right)^{2}-\frac{1}{r^2}\left(\frac{\partial S}{\partial \phi}\right)^{2}-\mu_{0}^{2}-2\alpha\frac{2C(r)}{r^{2}}\left(\frac{\partial S}{\partial r}\right)^{2}\left(\frac{\partial S}{\partial \phi}\right)^{2}
\nonumber \\
+2\alpha \left[\mu_{0}^{4}-\frac{1}{r^{4}}\left(\frac{\partial S}{\partial \phi}\right)^{4}-C(r)^{2}\left( \frac{\partial S}{\partial r}\right)^{4}\right]=0,
\label{EqnD2}
\end{eqnarray}
where the terms with higher order $\alpha$ parameter are neglected. Afterwards, using the Eq.(\ref{Radial}), the radial trajectory of the tunnelling Dirac particle, $W_{\pm}(r)$, is found as
\begin{equation}
W_{\pm }(r)=\pm \int \frac{\sqrt{E^2-C(r)\left(\mu_{0}^{2}+j^{2}/r^{2}\right)}}{C(r)}\left[1+\alpha \chi\right]dr, \label{EqnD3}
\end{equation}
where $\chi$ is
\begin{eqnarray*}
\chi=\frac{E^{2}(2\mu_{0}^{2}C(r)-E^{2})}{C(r)\left[E^{2}-C(r)\left(\mu_{0}^{2}+j^{2}/r^{2}\right)\right]}.
\end{eqnarray*}
Under the near trapping horizon condition, $C(r)\approx(r-r_{h})a$, $W_{\pm}(r)$ are obtained as
\begin{equation}
W_{\pm}(r_{h})=\pm i\pi \frac{E}{a}\left[1+\frac{\alpha}{2}(3\mu_{0}^{2}-\frac{j^{2}}{r_{h}^{2}})\right].\label{EqnD4}
\end{equation}
Using the Eqs.(\ref{Eqn1}) and (\ref{Eqn2}), the modified Hawking temperature of the Dirac particle, $T_{H}^{D}$ is obtained as follows:
\begin{equation}
T_{H}^{D}=\frac{T_{H}}{\left[1+\frac{\alpha}{2}(3\mu_{0}^{2}-\frac{j^{2}}{r_{h}^{2}})\right]}\approx\ T_{H}\left[1-\frac{\alpha}{2}(3\mu_{0}^{2}-\frac{j^{2}}{r_{h}^{2}})\right], \label{EqnD5}
\end{equation}
where $T_{H}$ is the usual Hawking temperature in Eq.(\ref{Eqn5}). This result indicates that the modified Hawking temperature of the Dirac particle is completely different from that of the scalar particle. In addition, it depends on the properties of the Dirac particle as well as the throat radius of the wormhole, as in Dirac particle tunneling from black holes and strings when the GUP effect is taken into account \cite{a1,a2,a3,a5,GY4,GY6,GY7,GY8,GY9,Kuntz,Casadio}. In the absence of the GUP effect, the modified Hawking temperature reduces to the standard one.

\subsection{Tunneling of massive vector boson from the wormhole}

The (2+1) dimensional modified massive vector boson equation is;
\begin{eqnarray}
i\beta^{0}(x)\partial_{0}\widetilde{\Psi}+i\beta^{i}(x)\left(1-\alpha m_{0}^{2}\right)\partial_{i}\widetilde{\Psi}+i\alpha
\hbar^{2}\beta^{i}(x)\partial_{i}\left(\partial_{j}\partial^{j}\widetilde{\Psi}\right)\nonumber \\-\frac{m_{0}}{\hbar}\left(1+\alpha\hbar^{2}\partial_{j}\partial^{j}-\alpha m_{0}^{2}\right)\widetilde{\Psi}-i\beta^{\mu}(x)\Sigma_{\mu}\left(1+\alpha\hbar^{2}\partial_{j}\partial^{j}-\alpha m_{0}^{2}\right)\widetilde{\Psi}=0,  \label{ModMVB}
\end{eqnarray}
where the $\widetilde{\Psi}$ and $m_{0}$ are the modified wave function and
mass of the vector boson, respectively \cite{GY5,GY8}. Also, the $\beta^{\mu}(x)$ and $\Sigma_{\mu}$ are
the Kemmer matrices and spin connection coefficients for the spin-1 given as
\begin{eqnarray}
\beta^{\mu}(x)=\overline{\sigma}^{\mu}(x)\otimes I+I\otimes \overline{\sigma}^{\mu}(x)\nonumber \\
\Sigma_{\mu}\left(x\right)=\Gamma_{\mu}(x)\otimes I+I\otimes\Gamma_{\mu}(x) \label{SpinCon}
\end{eqnarray}
respectively \cite{Sucu2,Sucu3}. To analyze the Hawking temperature of the wormhole in the framework of GUP via tunnelling of the spin-1 vector boson particle, we use the following ansatz for the wave function \cite{Sucu2,Sucu3};
\begin{eqnarray}
\widetilde{\Psi}(x)=\exp \left( \frac{i}{\hbar }S\left( t,r,\phi \right)
\right)\ \left(\begin{array}{c}A\left( t,r,\phi \right) \\B\left(
t,r,\phi \right) \\B\left(
t,r,\phi \right) \\D\left(
t,r,\phi \right)\\ \end{array}\right).  \label{ansatzMVB}
\end{eqnarray}
Then, using Eqs.(\ref{connec}), (\ref{SpinCon}) and (\ref{ansatzMVB}), the
modified massive vector boson equation is simplified to the three coupled
differential equations:
\begin{eqnarray*}
B\left[\alpha \frac{1}{r^{3}}\left( \frac{\partial S}{\partial \phi}\right)^{3}+i\alpha \frac{\sqrt{C(r)}}{r^{2}}\left(\frac{
\partial S}{\partial\phi}\right)^{2}\left(\frac{\partial S}{\partial r}\right)-i\alpha m_{0}^{2}\sqrt{C(r)}\frac{\partial S}{\partial r}+i\sqrt{C(r)}\frac{\partial S}{\partial r}\right] \\
+B\left[\frac{1}{r}\frac{\partial S}{\partial \phi }+\alpha\frac{C(r)}{r}\left(\frac{\partial S}{\partial\phi}\right)\left(\frac{\partial S}{\partial r}\right)^{2}+i\alpha C(r)\sqrt{C(r)}\left(\frac{\partial S}{\partial r}\right)^{3}-\alpha \frac{m_{0}^{2}}{r}\frac{\partial S}{\partial \phi}\right]\\ +A\left[\frac{\partial S}{\partial t}+\frac{m_{0}\left(1-\alpha m_{0}^{2}\right) }{2}+\alpha\frac{m_{0}}{2r^{2}}\left(\frac{\partial S}{\partial\phi}\right)^{2}+\alpha \frac{m_{0}C(r)}{2}\left(\frac{\partial S}{\partial r}\right)^{2}\right]=0
\end{eqnarray*}
\begin{eqnarray*}
A \left[\alpha \frac{C(r)}{r}\left(\frac{\partial S}{\partial \phi }\right) \left( \frac{\partial S}{\partial r}\right)^{2}-\alpha \frac{m_{0}^{2}}{r}\frac{\partial S}{\partial \phi}-i\alpha C(r)\sqrt{C(r)}\left(\frac{\partial S}{\partial r}\right)^{3}+\frac{\alpha}{r^{3}}\left(\frac{\partial S}{\partial \phi }\right)^{3}\right] \\
+A\left[-i\sqrt{C(r)}\frac{\partial S}{\partial r}+\frac{1}{r}\frac{\partial S}{\partial\phi}-i\alpha\frac{\sqrt{C(r)}}{r^{2}}\left(\frac{\partial S}{\partial\phi}\right)^{2}\left(\frac{\partial S}{\partial r}\right)+\alpha m_{0}^{2}\sqrt{C(r)}\frac{\partial S}{\partial r}\right]\\
+D\left[\alpha\frac{m_{0}^{2}}{r}\frac{\partial S}{\partial\phi}-\alpha \frac{C(r)}{r}\left(\frac{\partial S}{\partial\phi}\right)\left(
\frac{\partial S}{\partial r}\right)^{2}-i\sqrt{C(r)}\frac{\partial S}{\partial r}-i\alpha C(r)\sqrt{C(r)}\left(\frac{\partial S}{\partial r}\right)^{3}\right] \\
+D\left[-\frac{1}{r}\frac{\partial S}{\partial\phi}+i\alpha m_{0}^{2}\sqrt{C(r)}\frac{\partial S}{\partial r}-\frac{\alpha}{r^{3}}\left(\frac{\partial S}{\partial\phi}\right)^{3}-i\alpha\frac{\sqrt{C(r)}}{r^{2}}\left(\frac{\partial S}{\partial \phi}\right)^{2}\left(\frac{\partial S}{\partial r}\right)\right] \\+B\left[-m_{0}\left(1-\alpha m_{0}\right)-\alpha m_{0}C(r)\left(\frac{\partial S}{\partial r}\right)^{2}-\alpha \frac{m_{0}}{r^{2}}\left(\frac{\partial S}{\partial\phi}\right)^{2}\right]=0
\end{eqnarray*}
\begin{eqnarray}
B\left[-i\sqrt{C(r)}\frac{\partial S}{\partial r}-i\alpha \frac{\sqrt{C(r)}}{r^{2}}\left(\frac{\partial S}{\partial\phi}\right)^{2}\left(\frac{\partial S}{\partial r}\right)+i\alpha m_{0}^{2}\sqrt{C(r)}\frac{\partial S}{\partial r}+\frac{\alpha}{r^{3}}\left(\frac{\partial S}{\partial\phi}\right)^{3}\right]\nonumber \\ +B\left[\frac{1}{r}\frac{\partial S}{\partial\phi}+\alpha\frac{C(r)}{r}\left(\frac{\partial S}{\partial\phi}\right)\left(\frac{\partial S}{\partial r}\right)^{2}-\alpha\frac{m_{0}^{2}}{r}\frac{\partial S}{\partial\phi}-i\alpha C(r)\sqrt{C(r)}\left(\frac{\partial S}{\partial r}\right)^{3}\right]\nonumber \\
+D\left[\frac{\partial S}{\partial t}-\frac{m_{0}\left(1-\alpha m_{0}^{2}\right)}{2}-\alpha \frac{m_{0}}{2r^{2}}\left(\frac{\partial S}{\partial\phi}\right)^{2}-\alpha \frac{m_{0}C(r)}{2}\left(\frac{\partial S}{\partial r}\right)^{2}\right]=0  \label{EqnMVB1}
\end{eqnarray}
The Eq.(\ref{EqnMVB1}) has nontrivial solutions for the coefficients $A\left(t,r,\phi \right)$, $B\left( t,r,\phi \right)$ and $D\left( t,r,\phi\right)$ under the condition that the coefficients matrix determinant is zero.
Hence, the modified Hamilton-Jacobi equation for the massive vector boson particle becomes as follows;
\begin{eqnarray}
\left(\frac{\partial S}{\partial t}\right)^{2}-\frac{1}{r^{2}}
\left(\frac{\partial S}{\partial \phi }\right)^{2}-C(r)\left(\frac{\partial S}{\partial r}\right)^{2}-\frac{m_{0}^{2}}{4}+\alpha \left[\frac{9m_{0}^{2}C(r)}{4}\left(\frac{\partial S}{\partial r}
\right)^{2}+\frac{9m_{0}^{2}}{4r^{2}}\left(\frac{\partial S}{\partial\phi}\right)^{2}\right]\nonumber \\+\alpha\left[\frac{3m_{0}^{4}}{4}-\frac{6C(r)}{r^{2}}\left(\frac{\partial S}{\partial r}\right)^{2}\left(\frac{\partial S}{\partial\phi}\right)^{2}-3C(r)^{2}\left(\frac{\partial S}{\partial r}\right)^{4}-\frac{3}{r^{4}}\left(\frac{\partial S}{\partial \phi}
\right)^{4}\right]\nonumber \\+ \alpha \left[\frac{1}{r^{2}}\left(\frac{\partial S}{\partial t}
\right)^{2}\left(\frac{\partial S}{\partial \phi}\right)^{2}-m_{0}^{2}\left( \frac{\partial S}{\partial t}
\right)^{2}+C(r)\left(\frac{\partial S}{\partial t}\right)^{2}\left(\frac{\partial S}{\partial r}\right)^{2}\right]=0 \label{EqnMVB2}
\end{eqnarray}
Using the method of separating the variable of the classical action function, as in Eq.(\ref{Radial}), the radial trajectory, $W_{\pm}\left(r\right)$, of the massive vector boson is written as follows:
\begin{eqnarray}
W_{\pm}(r)=\pm \int \frac{\sqrt{E^2-C(r)\left(m_{0}^{2}+j^{2}/r^{2}\right)}}{C(r)}\left[1+\alpha\Upsilon\right]dr \label{EqnMVB3}
\end{eqnarray}
where $\Upsilon$ is
\begin{eqnarray*}
\Upsilon=\frac{E^{2}(5m_{0}^{2}C(r)-E^{2})}{4C(r)\left[E^{2}-C(r)\left(m_{0}^{2}+j^{2}/r^{2}\right)\right]}.
\end{eqnarray*}
Using the near trapping horizon condition, $C(r)\approx(r-r_{h})a$, the integral expression in Eq.(\ref{EqnMVB3}) are calculated as
\begin{eqnarray}
W_{\pm}(r_{h})=\pm i\pi \frac{E}{a}\left[1+\frac{\alpha}{2}(\frac{9}{4}m_{0}^{2}-\frac{j^{2}}{r_{h}^{2}})\right]. \label{EqnMVB4}
\end{eqnarray}
Hence, using the Eqs.(\ref{Eqn1}) and (\ref{Eqn2}) for the vector
boson particle, the modified Hawking temperature becomes
\begin{eqnarray}
T_{H}^{^{\prime}}=\frac{T_{H}}{\left[1+\frac{\alpha}{2}(\frac{9}{4}m_{0}^{2}-\frac{j^{2}}{r_{h}^{2}})\right]}\approx T_{H}\left[1-\frac{\alpha}{2}(\frac{9}{4}m_{0}^{2}-\frac{j^{2}}{r_{h}^{2}})\right].\label{EqnMVB5}
\end{eqnarray}
Here, $T_{H}$ is the usual Hawking temperature of the wormhole. As can be seen from Eq.(\ref{EqnMVB5}), the modified Hawking temperature of the vector boson different from that of the both Dirac and scalar particles. And also, as a result of the quantum gravity, the modified Hawking temperature depends on the mass and the angular momentum of the vector boson particle, as in vector boson particle tunneling from a black hole in the presence of the GUP effect \cite{GWS1,GWS2,GWS3,GY5,GY8,GY9}.

\section{Concluding remarks}\label{conc}
In this work, in the framework of the quantum gravity, we have investigated Hawking temperature of the $(2+1)$ dimensional traversable wormhole by using the quantum tunnelling processes of the spin-0 scalar, spin-1/2 Dirac and spin-1 vector boson particles, respectively. Some important results deduced from this study can be listed as follows:

\begin{itemize}
\item We observe that due to the presence of a trapping horizon, the Hamilton-Jacobi approach is found to be useful in determining the thermal properties of wormholes.
\item As can be seen from Eq.(\ref{Eqn5}), the standard Hawking temperature of the $(2+1)$ dimensional traversable wormhole is positively defined and depends only on the wormhole properties. The positive temperature indicates that a wormhole can be treated as a physical system. This indicates that it may be supported by an ordinary (i.e. non-exotic) matter.
\item The Unruh-Verlinde temperature of the wormhole was calculated by using the vector Kodama vector instead of the time-like Killing vector. It is observed that, at throat of the wormhole, the Unruh-Verlinde temperature is equal to the standard Hawking temperature. Also, both temperatures are non-vanishing. Accordingly, it can be said that \textit{Kodama} and \textit{Killing observers} detected a different temperatures in the $(2+1)$ dimensional traversable wormhole background.
\item This situation indicates that the  reason why Kodama and Killing observers measure different temperatures is directly related to the geometric point of view.
\item As a result in quantum gravity effect, the modified Hawking temperatures caused by the tunneling of the spin-0 scalar, spin-1/2 Dirac and spin-1 particles, separately, are lower than the standard one, and also the modified Hawking temperature depends on both the properties of wormhole and tunnelled particles. Moreover, for all the three types of particles, angular momentum (orbital+spin)-spacetime geometry interaction appears as $\frac{j^2}{r_{h}^{2}}$ in the context of the GUP, where $\textit{j}$ is the angular momentum (orbital+spin) of tunneling particle and $r_{h}$ is radius of the wormhole throat.
\item In the context of GUP, of course, the tunneling probability of the spinning particles from the wormhole is different from each other.
\end{itemize}

Finally, as in black holes, similarly, it can be said that the quantum mechanical tunnelling processes of the relativistic spinning particles play an important role in the investigation of the thermodynamic properties of a wormhole in the presence of the quantum gravity effects.

\section*{Conflict of Interests}
The authors declare that there is no conflict of interests regarding the publication of this paper.

\section*{Data Availability}
No data were used to support this study.

\section*{Acknowledgements}
This work was supported by Research Fund of the Akdeniz University (Project No: FDK-2017-2867)
and the Scientific and Technological Research Council of Turkey (TUBITAK Project No: 116F329).

\end{document}